\documentclass[twocolumn, aps, prl, nofootinbib,superscriptaddress]{revtex4}
\usepackage{graphicx}
\usepackage{amsfonts}
\usepackage{amssymb}
\usepackage{amsthm}
\usepackage{array}
\usepackage{amsmath}
\usepackage{verbatim} 
\usepackage[colorlinks, linkcolor=blue, anchorcolor=blue, citecolor=blue, urlcolor=blue]{hyperref}
\usepackage{color}
\usepackage{bbold}
\usepackage{epstopdf}
\usepackage{mathtools}

\newcommand{\ket}[1]{\left\vert#1\right\rangle}
\newcommand{\bra}[1]{\left\langle#1\right\vert}

\def\bra#1{\langle #1|}
\def\ket#1{|#1 \rangle}

\def\Tr{\mbox{Tr}}

\begin{document}
\title{Demonstrating Quantum Microscopic Reversibility Using Coherent States of Light}
\author{Marco Bellini}
\email{marco.bellini@ino.cnr.it}
\affiliation{Istituto Nazionale di Ottica (CNR-INO), Largo E. Fermi 6, 50125 Florence, Italy}
\affiliation{LENS and Department of Physics and Astronomy, University of Firenze, 50019 Sesto Fiorentino, Florence, Italy}
\author{Hyukjoon Kwon}
\email{hjkwon@kias.re.kr}
\affiliation{Korea Institute for Advanced Study, Seoul 02455, South Korea}
\author{Nicola Biagi}
\affiliation{Istituto Nazionale di Ottica (CNR-INO), Largo E. Fermi 6, 50125 Florence, Italy}
\affiliation{LENS and Department of Physics and Astronomy, University of Firenze, 50019 Sesto Fiorentino, Florence, Italy}
\author{Saverio Francesconi}
\affiliation{Istituto Nazionale di Ottica (CNR-INO), Largo E. Fermi 6, 50125 Florence, Italy}
\affiliation{LENS and Department of Physics and Astronomy, University of Firenze, 50019 Sesto Fiorentino, Florence, Italy}
\author{Alessandro Zavatta}
\affiliation{Istituto Nazionale di Ottica (CNR-INO), Largo E. Fermi 6, 50125 Florence, Italy}
\affiliation{LENS and Department of Physics and Astronomy, University of Firenze, 50019 Sesto Fiorentino, Florence, Italy}
\author{M. S. Kim}
\email{m.kim@imperial.ac.uk}
\affiliation{QOLS, Blackett Laboratory, Imperial College London, London SW7 2AZ, United Kingdom}

\begin{abstract}
The principle of microscopic reversibility lies at the core of fluctuation theorems, which have extended our understanding of the second law of thermodynamics to the statistical level. In the quantum regime, however, this elementary principle should be amended as the system energy cannot be sharply determined at a given quantum phase space point. In this Letter, we propose and experimentally test a quantum generalization of the microscopic reversibility when a quantum system interacts with a heat bath through energy-preserving unitary dynamics. Quantum effects can be identified by noting that the backward process is less likely to happen in the existence of quantum coherence between the system's energy eigenstates. The experimental demonstration has been realized by mixing coherent and thermal states in a beam-splitter, followed by heterodyne detection in an optical setup. We verify that the quantum modification for the principle of microscopic reversibility is critical in the low-temperature limit, while the quantum-to-classical transition is observed as the temperature of the thermal field gets higher.
\end{abstract}
\pacs{}
\maketitle
\noindent
\noindent

Fluctuation theorems \cite{Gallavotti95, Jarzynski97, Crooks99} set the cornerstone for describing the statistical properties of non-equilibrium thermodynamics. Throughout the development of fluctuation theorems, the principle of microscopic reversibility has served as the main pillar, by providing a universal symmetry relation between a phase space trajectory and its time-reversed version. The second law of thermodynamics naturally arises from averaging fluctuating quantities over all possible trajectories, which leads to macroscopic irreversibility. Fluctuation theorems have been experimentally verified for various types of systems \cite{Liphardt02, Collin05, Blickle06, Saira12} in the classical regime, where the thermodynamic elements, such as work, energy, and heat, are determined from the phase space trajectories by well-established classical definitions.

On the other hand, there exists a fundamental difference between quantum and classical phase space structures, as the uncertainty principle prohibits the position and momentum of a quantum system from being measured precisely at the same time. Furthermore, energy is not necessarily a unique basis to investigate the thermodynamic quantities of a quantum state and its evolution. A quantum state can be prepared in a superposition of energy eigenstates, and measurement can be performed for an observable that does not commute with the system Hamiltonian. Considerable efforts from both thermodynamics \cite{Esposito09, Campisi11, Deffner11, Deffner13, Hanggi15, Jarzynski15, Funo18, Taranto18} and quantum information \cite{Goold16, Alhambra16, Binder19, Landi21} perspectives have been put into an investigation of how quantum coherence changes the fluctuation theorems, as well as how we generalize thermodynamic quantities, such as work and heat, consistent with quantum mechanics \cite{Talkner07, Bera19}. Throughout these efforts, quantum optics has played an important role by providing a promising platform to simulate quantum processes in the microscopic regime \cite{Scully03, Pikovski12, Vidrighin16}.

Recent developments of quantum fluctuation theorems \cite{Albash13, Batalhao14, Aberg18, Manzano18, Holmes19, Kwon19, Micadei20, Micadei21, Goold21, Yada22} have proposed that the symmetry relation between the forward and backward quantum thermodynamic processes should be modified properly to fully describe the quantum trajectories including the contributions from coherence. In this Letter, we establish a method to explore the difference between quantum and classical theory of microscopic reversibility on a very basic level and demonstrate it using a quantum optical setup. In the experiment, an initial coherent state of light interacts with a thermal field, and the resulting mixed state is then measured on a coherent basis. The measured statistics clearly follow the quantum version of microscopic reversibility, and we identify that quantum coherence plays a crucial role in the low-temperature limit. The experimental data agree with the theoretically predicted behaviors of the quantum modification with respect to the degree of coherence and the temperature of the thermal field.

{\it Microscopic reversibility in classical dynamics.---}
Suppose that a physical system is interacting with a thermal bath at temperature $T$. In classical statistical mechanics, a phase space trajectory of the dynamics is characterized by the system's position and momentum at each time $t$ as $z(t) = (x(t),p(t))$. Consequently, the time-reversed trajectory is defined as $\bar{z}(-t) = (\bar{x}(-t)),\bar{p}(-t)) = (x(-t),-p(-t))$ by the transformation $\bar{x} = x$ and $\bar{p} = -p$. The principle of microscopic reversibility tells us that there exists a symmetry relation between the probabilities of the trajectory $z(t)$ following the time-forward dynamics and the trajectory $\bar{z}(-t)$ following the time-reversed dynamics. When considering only the initial and final phase space points $z_i =(x_i,p_i)$ and $z_f = (x_f,p_f)$, the symmetry relation between the forward transition probability ${\cal P}_\rightarrow \left(z_f | z_i \right)$ from $z_i$ to $z_f$ and the backward transition probability ${\cal P}_\leftarrow \left(\bar{z}_i | \bar{z}_f \right)$ from $\bar{z}_f$ to $\bar{z}_i$ can be simplified as
\begin{equation}
\label{Eq:ClassicalDB}
\frac{{\cal P}_\rightarrow \left(z_f | z_i \right)}{{\cal P}_\leftarrow \left(\bar{z}_i| \bar{z}_f \right)} = e^{-\beta Q},
\end{equation}
where $Q$ is the heat flow from the bath to the system during the forward process and $\beta = 1/(k_B T)$ with the Boltzmann constant $k_B$. If the system-bath is isolated and no work has been done to the system, the amount of heat is equal to the system's energy change $ Q = \Delta E = H_f(z_f) - H_i(z_i)$, where $H_{i}(z_i)$ and $H_{f}(z_f)$ denote the system's Hamiltonians at initial and final points. From the microscopic reversibility condition in Eq.~\eqref{Eq:ClassicalDB}, various types of classical fluctuation theorems, including the Jarzynski equality \cite{Jarzynski97} and Crooks fluctuation theorem \cite{Crooks99} can be derived.

\begin{figure}[t]
\includegraphics[width=1.00\linewidth]{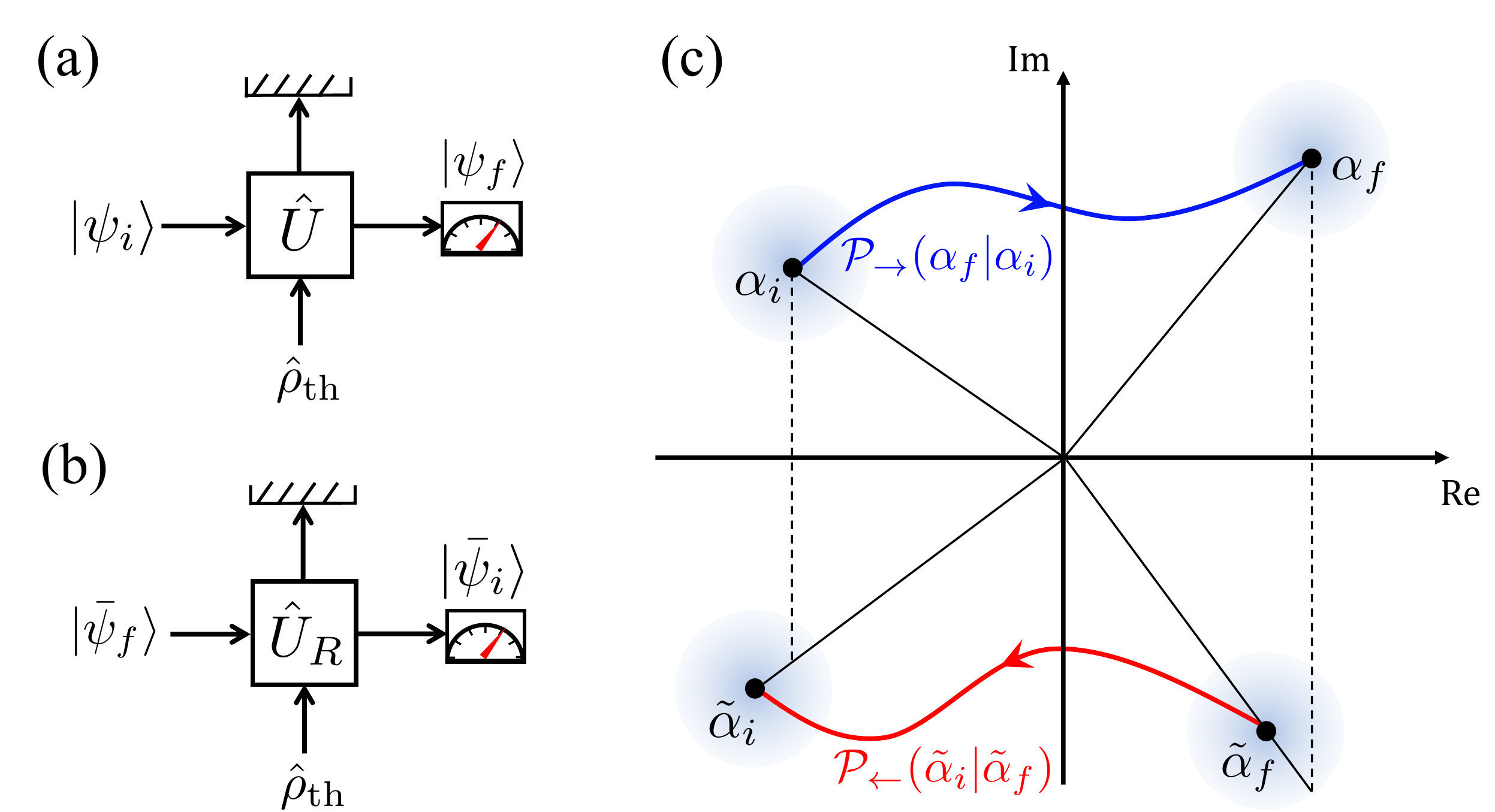}
\caption{(a) Forward and (b) backward quantum transition probabilities when a quantum state is interacting with the bath $\hat{\rho}_{\rm th}$ and (c) the rescaling of time-reversed coherent states.}
\label{fig:scheme}
\end{figure}

{\it Microscopic reversibility in quantum dynamics.---} In quantum mechanics, however, phase space should be defined differently as position and momentum are promoted to operators $\hat{x}$ and $\hat{p}$, which do not commute to each other, i.e., $[\hat{x}, \hat{p}] = i\hbar$. By taking into account the non-classical nature of quantum mechanics, we can construct a phase space structure for a quantum harmonic oscillator system with the Hamiltonian $\hat{H}(\hat{x},\hat{p}) = \frac{\hbar \omega_0}{2} \left( \hat{x}^2 + \hat{p}^2 \right) = \hbar \omega_0 \left( \hat{a}^\dagger \hat{a} + \frac{1}{2} \right)$, where $\hat{a} = (\hat{x} + i\hat{p})/\sqrt{2}$ is a bosonic annihilation operator. In particular, coherent states, satisfying $\hat{a} \ket{\alpha} = \alpha \ket{\alpha}$, compose a family of the most classical states among pure quantum states \cite{Mandel86, Zurek93}. The coherent state saturates the minimum uncertainty relation, and its distribution is centered at  $z = (x,p) = ({\rm Re} [\alpha], {\rm Im}[\alpha])/\sqrt{2}$ in quantum phase space. While the coherent state resembles a classical state at a certain phase space point, an important difference arises from the fact that the state contains quantum coherence between different energy states $\ket{n}$; $\ket{\alpha} = e^{-|\alpha|^2/2} \sum_{n=0}^\infty \left(\alpha^n/\sqrt{n!}\right) \ket{n}$.

Now let us generalize the classical scenario to the quantum harmonic oscillator system by interacting the system with a bath in thermal equilibrium, $\hat{\rho}_{\rm th} = e^{-\beta \hat{H}_B}/\Tr[e^{-\beta \hat{H}_B}]$, where $\hat{H}_B$ is a bath Hamiltonian. By denoting the unitary operator generated by an interaction Hamiltonian $\hat{H}_I$ as $\hat{U} = e^{- i \tau \hat{H}_I}$, the transition probability from the initial state $\ket{\psi_i}$ to the final state $\ket{\psi_f}$ is defined as 
\begin{equation}
{\cal P}_\rightarrow(\psi_f | \psi_i) = {\Tr} \left[ \hat{U}(\ket{\psi_i} \bra{\psi_i} \otimes \hat{\rho}_{\rm th}) \hat{U}^\dagger \left( \ket{\psi_f} \bra{\psi_f} \otimes \mathbb{\hat{1}}_B \right) \right].
\end{equation}

The quantum mechanical extension of the microscopic reversibility condition can be established by finding the backward process that corresponds to a generalized time-reversal~\cite{Aberg18, Kwon19}. In particular, when the system's equilibrium state $\hat\rho_{\rm eq} = e^{-\beta \hat{H}}/{\rm Tr} [ e^{-\beta \hat{H}} ]$ remains unchanged by the system-bath interaction, the following relation holds \cite{Kwon19} between the forward and backward probabilities:
\begin{equation}
\frac{{\cal P}_\rightarrow(\psi_f | \psi_i)}{{\cal P}_\leftarrow( \tilde\psi_i | \tilde\psi_f)} 
= \bra{\psi_i} e^{\beta \hat{H}} \ket{\psi_i} \bra{\psi_f} e^{-\beta \hat{H}} \ket{\psi_f},
\label{Eq:QuantumDB_general}
\end{equation}
where $\ket{\tilde\psi_i} \propto e^{\beta \hat{H}/2 }\ket{\bar\psi_i}$ and $\ket{\tilde\psi_f} \propto e^{-\beta\hat{H}/2} \ket{\bar\psi_f}$ with the normalization $\langle \tilde\psi_{i(f)} | \tilde\psi_{i(f)} \rangle = 1$. Here, $\ket{\bar\psi}$ denotes the time-reversed state \cite{footnote} of $\ket{\psi}$.

However, the explicit construction of the backward process in Eq.~\eqref{Eq:QuantumDB_general} can be very complicated in general. To consider the simplest scenario, we assume that the system and bath Hamiltonians are time-independent and invariant under time-reversal, i.e., $\hat{H}(\hat{x}, \hat{p}) = \hat{H}(\hat{x}, -\hat{p})$ and $\hat{H}_B(\hat{x}_B, \hat{p}_B) = \hat{H}_B (\hat{x}_B, -\hat{p}_B)$, and the interaction Hamiltonian satisfies the energy conservation condition,
\begin{equation}
[\hat{H}_{I}, \hat{H} + \hat{H}_B ] =0.
\label{Eq:energy_conservation}
\end{equation}
We note that the equilibrium state $\hat\rho_{\rm eq}$ becomes a fixed point under the dynamics as $\hat{U} (\hat\rho_{\rm eq} \otimes \hat\rho_{\rm th}) \hat{U}^\dagger = \hat\rho_{\rm eq} \otimes \hat\rho_{\rm th}$. Under these assumptions, the backward dynamics is simply described by the time-reversed unitary operator $\hat{U}_R = e^{-i \tau \hat{H}_{I}^R }$, where $\hat{H}_{I}^R(\hat{x}, \hat{p}, \hat{x}_B, \hat{p}_B) = \hat{H}_{I}(\hat{x}, -\hat{p}, \hat{x}_B, -\hat{p}_B)$ \cite{Aberg18}. Consequently, the backward probability of the quantum trajectory becomes
\begin{equation}
{\cal P}_\leftarrow(\bar\psi_i | \bar\psi_f) = {\Tr} \left[ \hat{U}_R (\ket{\bar\psi_f} \bra{\bar\psi_f} \otimes \hat{\rho}_{\rm th}) \hat U_R^\dagger \left( \ket{\bar\psi_i} \bra{\bar\psi_i} \otimes \mathbb{\hat{1}}_B \right) \right]
\end{equation}
 (see Fig.~\ref{fig:scheme}(a) and (b)).
For such a process, often referred to as a thermal operation \cite{Janzing00, Morodecki13, Lostaglio18}, the heat can be solely characterized by the system's energy change $Q = \Delta E$ so that no work is done on the system, just as in the previously discussed classical scenario. 

One may then ask whether this relation reduces to the classical microscopic reversibility condition in Eq.~\eqref{Eq:ClassicalDB} by taking $\ket{\psi_{i(f)}}= \ket{\alpha_{i(f)}}$ for the transitions between two coherent states. We note that this is not the case because coherent states are not the eigenstates of $\hat{H}$. Instead, from Eq.~\eqref{Eq:QuantumDB_general}, we find the microscopic reversibility condition between the coherent states,
\begin{equation}
\frac{{\cal P}_\rightarrow(\alpha_f | \alpha_i)}{{\cal P}_\leftarrow( \tilde\alpha_i | \tilde\alpha_f)} 
= \exp\left[-\frac{|\alpha_f|^2}{n_{\rm th} +1} + \frac{|\alpha_i|^2}{n_{\rm th}}\right],
%= \Upsilon e^{- \beta Q},
\label{Eq:QuantumDB}
\end{equation}
where $\tilde\alpha_i = \bar\alpha_i e^{\beta \hbar \omega_0/2}$ and $\tilde\alpha_f =  \bar\alpha_f e^{-\beta \hbar \omega_0/2}$ with $\bar\alpha$ being the complex conjugate of $\alpha$, and $n_{\rm th} = e^{-\beta \hbar \omega_0} / (1 - e^{-\beta \hbar \omega_0}) $ is the mean excitation number when the system equilibrates to the bath. We note that the ratio of the forward-to-backward transition probabilities is solely determined by the initial and final coherent state amplitudes.

We observe that two kinds of modifications are made for the microscopic reversibility condition in the quantum regime. The first comes from the rescaling of the reverse trajectory with respect to the bath temperature such that $|\tilde\alpha_i| \geq |\alpha_i|$ and $|\tilde\alpha_f| \leq |\alpha_f|$ (see Fig.~\ref{fig:scheme}(c)). This is because a quantum state can contain coherence between energy eigenstates, where each of them should be weighted differently depending on the energy and the bath temperature \cite{Aberg18, Kwon19}. While the rescaled states provide a quadratic correction in $\beta \hbar \omega_0$ by noting that $\langle \alpha | \alpha e^{\pm \beta \hbar \omega _0/2} \rangle \approx 1 - (|\alpha|^2/8) (\beta \hbar \omega_0)^2$, this so-called Gibbs rescaling is essential to establish the microscopic reversibility, such that the ratio between the forward and backward trajectories does not depend on the detailed form of the interaction unitary $\hat{U}$.

The second quantum modification can be captured by the following quantity,
\begin{equation}
\Upsilon = e^{\beta Q} \left[ \frac{{\cal P}_\rightarrow(\psi_f | \psi_i)}{{\cal P}_\leftarrow(\tilde\psi_i | \tilde\psi_f)} \right] = \frac{\bra{\psi_i}e^{\beta \hat{H}} \ket{\psi_i}}{e^{\beta \bra{\psi_i} \hat{H}\ket{\psi_i}}}\frac{\bra{\psi_f}e^{-\beta \hat{H}} \ket{\psi_f}}{e^{-\beta \bra{\psi_f}\hat{H}\ket{\psi_f}}},
\label{Eq:Modification_theory}
\end{equation}
where $Q = \Delta E = \bra {\psi_f} \hat{H} \ket{\psi_f} - \bra {\psi_i} \hat{H} \ket{\psi_i}$. The quantity $\Upsilon$ indicates how far the right-hand side of Eq.~\eqref{Eq:QuantumDB} deviates from $e^{-\beta Q}$. The last expression in Eq.~\eqref{Eq:Modification_theory} shows that the correction comes from the fact that $\bra\psi e^{\pm \beta \hat H} \ket\psi$ is always larger than $e^{\pm \beta \bra\psi \hat{H} \ket\psi}$ when a quantum state $\ket\psi$ contains superposition between energy eigenstates. Thus, the quantum effect always gives a correction in the way that the backward-to-forward process ratio is smaller than the classical prediction, i.e., $\Upsilon \geq 1$, which means that the Gibbs-rescaled backward process is less likely to happen if quantum coherence is involved. We find the explicit form of the quantum modification for coherent states, 
\begin{equation}
\begin{aligned}
 \log \Upsilon &= A |\alpha|^2_{\rm tot} - B \left( \Delta |\alpha|^2 \right),
 \end{aligned}
 \label{Eq:Modification_Pred}
\end{equation}
where $|\alpha|^2_{\rm tot} = |\alpha_i|^2 + |\alpha_f|^2$ and $\Delta |\alpha|^2 = |\alpha_f|^2 - |\alpha_i|^2$ with coefficients $A = \cosh(\beta \hbar \omega_0) -1$ and $ B = \sinh(\beta \hbar \omega_0) -\beta\hbar\omega_0$. The leading order of the quantum modification in $\Upsilon$ scales quadratically with respect to $\beta \hbar \omega_0$ and linearly with respect to $|\alpha|^2_{\rm tot}$. The larger the $|\alpha|^2$, the more different energy terms are involved in $\ket{\alpha}$ so the larger the coherence. More precisely, contribution of different energy terms in a coherent state $\ket\alpha$ can be quantified by the quantum Fisher information \cite{Kwon18}, which is proportional to the variance of energy, ${\rm Var}_{\ket{\alpha}}(\hat{H}) =  \bra{\alpha} \hat{H}^2 \ket{\alpha} - \bra{\alpha} \hat{H} \ket{\alpha}^2 = (\hbar \omega_0)^2 |\alpha|^2$. For a general transition from the state $\ket{\psi_i}$ to $\ket{\psi_f}$, the lowest order correction is given as $\log \Upsilon = \frac{\beta^2}{2}\left[{\rm Var}_{\ket{\psi_i}}(\hat{H}) + {\rm Var}_{\ket{\psi_f}}(\hat{H})\right] + {\cal O}((\beta\hbar\omega_0)^3)$. This implies that the quantum correction becomes even larger for the transition between non-classical states with high energy variances, such as squeezed states.

{\it Experimental test of microscopic reversibility.---}
\begin{figure}[t]
\includegraphics[width=.9\linewidth]{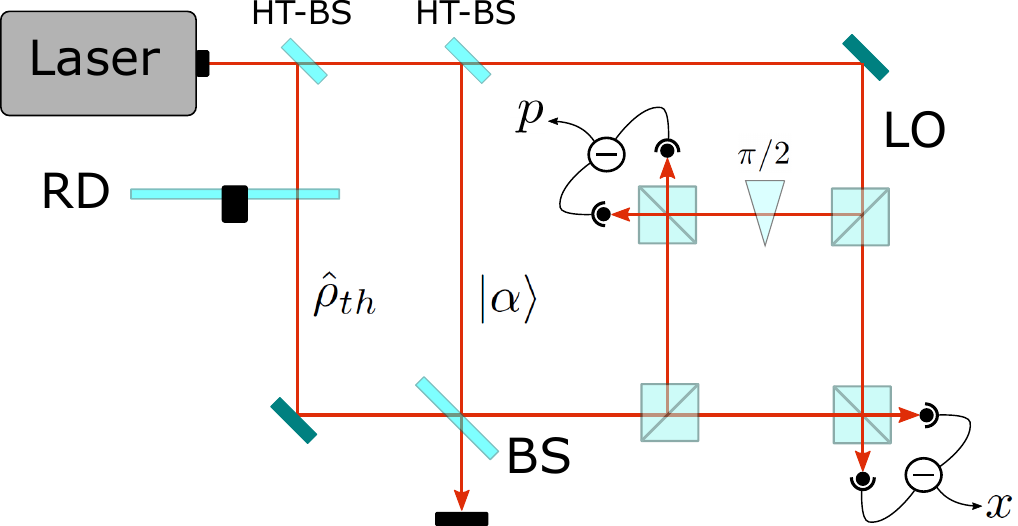}
\caption{Simplified experimental setup for testing the quantum fluctuation theorem. Cubes represent 50$\%$ beam-splitters, and HT-BS are high-transmittivity beam-splitters. All other symbols are defined in the text.
}
\label{Fig:ExpScheme}
\end{figure}
\begin{figure}[t]
\includegraphics[width=.9\linewidth]{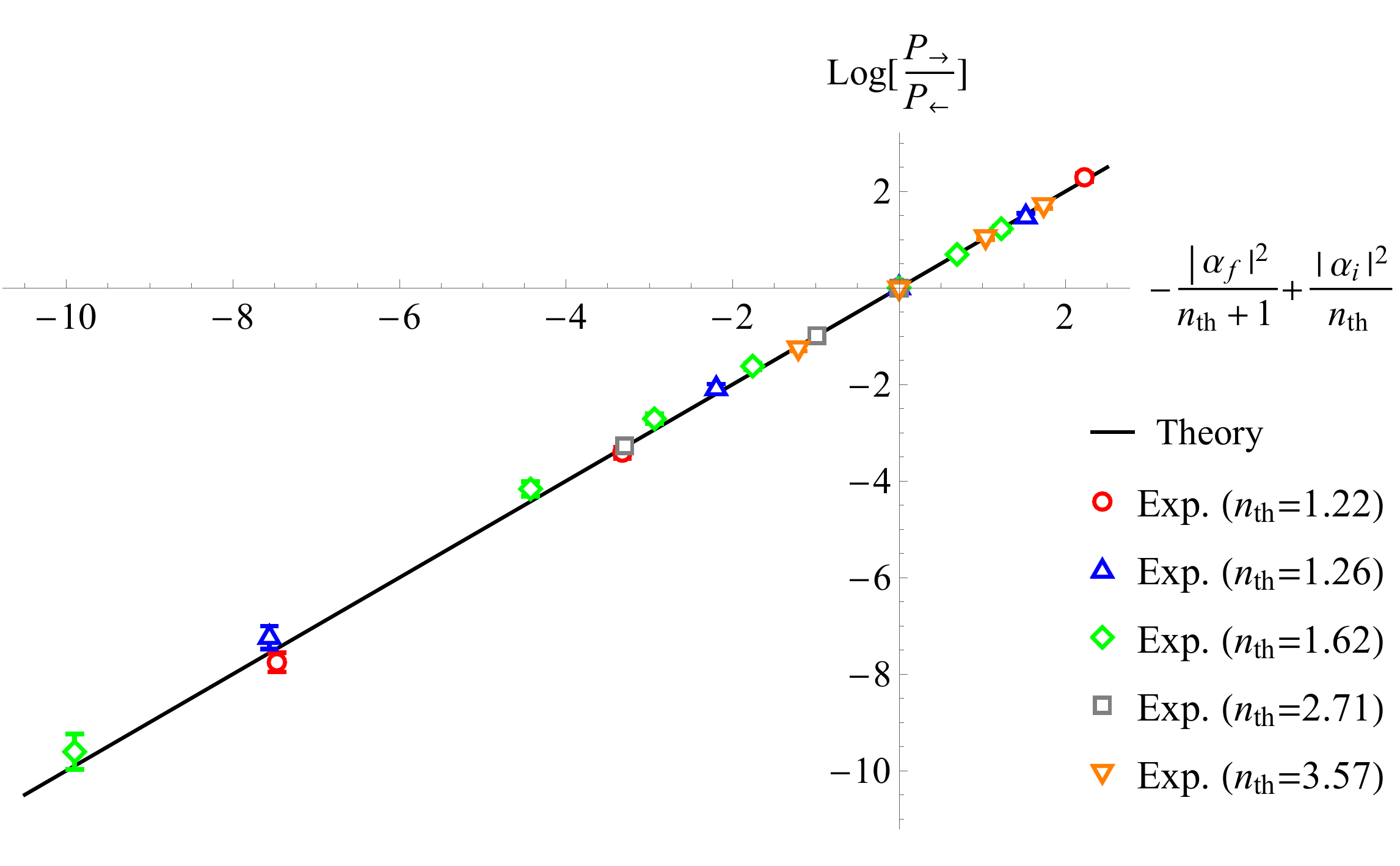}
\caption{Experimental data for the quantum microscopic reversibility described in Eq.~\eqref{Eq:QuantumDB}. We plot the logarithmic ratio of forward and backward transition probabilities using different initial state amplitudes ($|\alpha_i|$ in the range from 1.46 to 3.36) and different mean photon number thermal states ($n_{\rm th}$ in the range from 1.22 to 3.57). The reflectivity of the beam-splitter combining the coherent and thermal states is varied as well (85$\%$ for data with $n_{\rm th}=1.62$, 70$\%$ for the others). }
\label{Fig:ExpResults}
\end{figure}
We experimentally demonstrate the coherence effects on the microscopic reversibility using quantum optics. We take an optical mode $\hat{a}$ as a system and another optical mode $\hat{b}$ as a bath state, whose Hamiltonians are expressed as $\hat H = \hbar \omega_0 \left( \hat{a}^\dagger \hat{a} + \frac{1}{2} \right)$ and $\hat{H}_B = \hbar \omega_0 \left( \hat{b}^\dagger \hat{b} + \frac{1}{2} \right) = \frac{\hbar \omega_0}{2} (\hat{x}_B^2 + \hat{p}_B^2)$, respectively. The energy conservation condition in Eq.~\eqref{Eq:energy_conservation} of the interaction unitary can be achieved by interacting the system and bath through passive linear optical elements, which preserves the total photon number $\hat{N} = \hat{a}^\dagger \hat{a} + \hat{b}^\dagger \hat{b}$. In the experiment, we adopt the simplest model of linear system-bath interaction given by a beam-splitter $\hat{U}(\theta) = e^{ - i \theta (\hat{a} \hat{b}^\dagger + \hat{a}^\dagger \hat{b})} = e^{-i \theta (\hat{x}\hat{x}_B + \hat{p}\hat{p}_B) }$ with real values of $\theta$. By noting that the interaction Hamiltonian $\hat{H}_{I}(\hat{x},\hat{p},\hat{x}_B,\hat{p}_B) = \hat{x}\hat{x}_B + \hat{p}\hat{p}_B$ is invariant under the time-reversal $(\hat{x}, \hat{p}, \hat{x}_B, \hat{p}_B) \rightarrow (\hat{x}, -\hat{p}, \hat{x}_B, -\hat{p}_B)$, the backward unitary becomes the same as the forward unitary, i.e., $U_R(\theta) = U(\theta)$, which simplifies the experimental procedure.

The transition probability of the forward (backward) trajectory is obtained by preparing the coherent state $\ket {\alpha_i}$ $(\ket {\tilde \alpha_f})$ and interacting with the bath using the beam-splitter, followed by heterodyne measurements of the system state. The distribution of the heterodyne detection outcomes becomes the Husimi $Q$-function \cite{Richter98}, and the probability density at the point $\ket {\alpha_f}$ $(\ket {\tilde \alpha_i})$ indicates the transition probability of the forward (backward) trajectory by noting that the $Q$-function of a quantum state $\hat\rho$ is given by $Q_{\hat\rho} (\alpha) := \frac{1}{\pi} \bra\alpha \hat\rho \ket\alpha$.

Figure~\ref{Fig:ExpScheme} shows a simplified scheme of the experiment to test the microscopic reversibility condition in Eq.~\eqref{Eq:QuantumDB}. A pulsed Ti:sapphire laser is used as the main source to produce coherent states $\ket{\alpha}$ of adjustable amplitude, and thermal states $\hat{\rho}_{\rm th}$ of different mean photon number $n_{\rm th}$ by phase/amplitude randomization via a rotating glass disk (RD) \cite{Arecchi1965,Zavatta2007,Parigi2009}. After mixing them in a variable-ratio beam-splitter (BS), the output state undergoes heterodyne detection, using another portion of the laser emission as the local oscillator (LO), for the simultaneous measurement of its orthogonal field quadratures $\hat{x}$ and $\hat{p}$ and the reconstruction of its $Q$-function (more details on the experimental setup are available in the Supplemental Material \cite{Suppl}). 
The measurement procedure starts by fixing a mean photon number $n_{\rm th}$ of the thermal state, which is mixed with an initial coherent state $\lvert \alpha_i \rangle$ in the beam-splitter BS set at the appropriate value of reflectivity. Then, one measures the $Q$-function of the output state and, by evaluating it in $\alpha_f$, retrieves the transition probability of the forward process, ${\cal P}_\rightarrow(\alpha_f | \alpha_i)$. For the time-reversed process, we inject a coherent state with a Gibbs rescaled amplitude $\tilde\alpha_f = \bar\alpha_f e^{-\beta\hbar\omega_0/2}$ into the BS, set at the same reflectivity value and with the same thermal state at the other input. The $Q$-function of the output state is then finally evaluated in the Gibbs rescaled amplitude $\tilde \alpha_i = \bar\alpha_i e^{\beta\hbar\omega_0/2}$ to obtain ${\cal P}_\leftarrow( \tilde\alpha_i | \tilde\alpha_f)$.

Experimental results presented in Fig.~\ref{Fig:ExpResults} clearly demonstrate that the ratio between the forward and backward transition probabilities well matches the quantum prediction of the microscopic reversibility in Eq.~\eqref{Eq:QuantumDB} for a wide range of values of coherent state amplitudes and bath temperatures. Note that we used both real and complex values for the coherent state amplitudes $\alpha_i$. In the first case, the rescaling for the reverse process simply implies a change in the absolute value of the coherent state amplitudes, whereas in the second case it also involves a phase reversal due to complex conjugation. We also successfully verify that the microscopic reversibility condition holds regardless of the values of the beam-splitter reflectivity.

\begin{figure}[t]
\includegraphics[width=.9\linewidth]{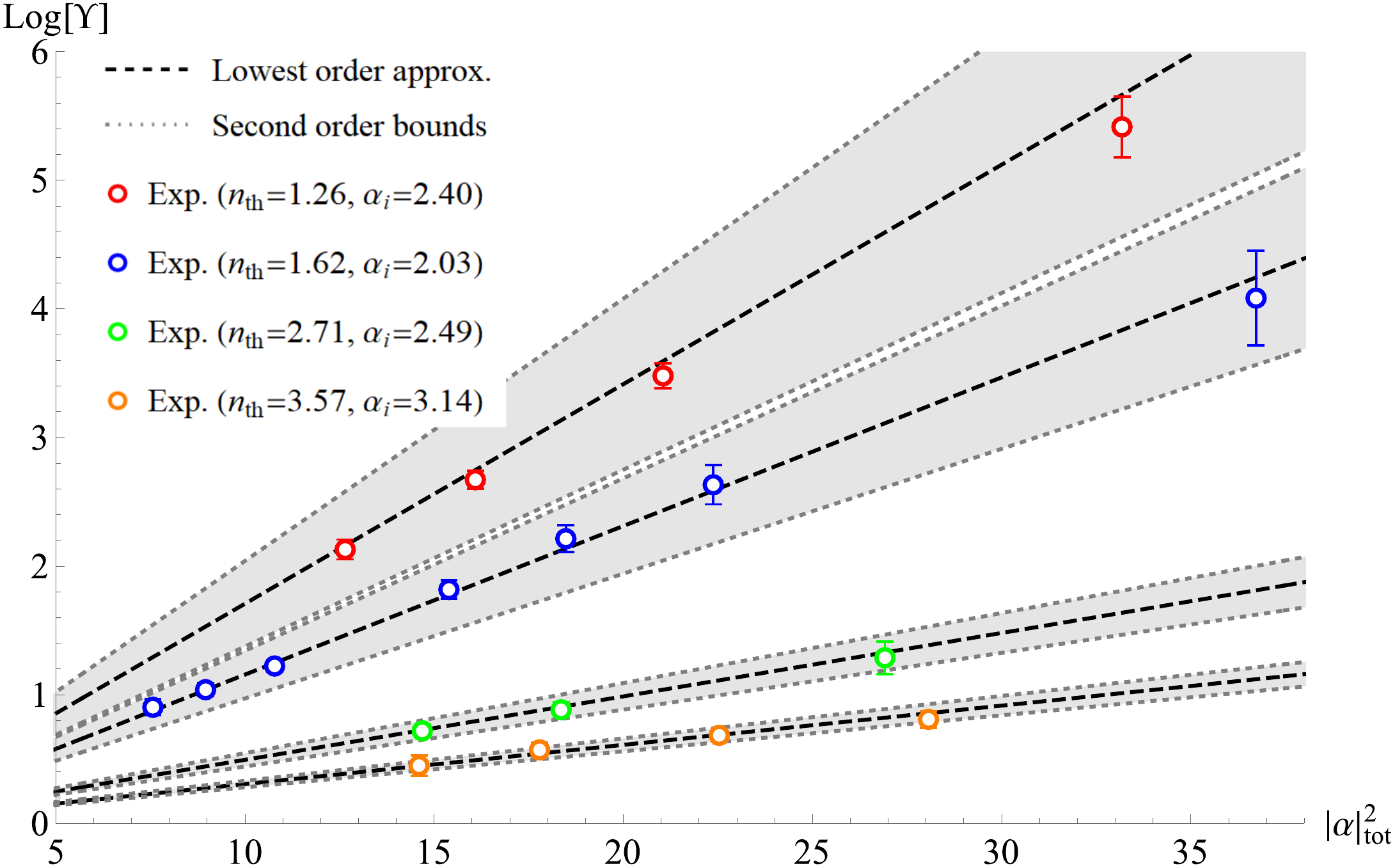}
\caption{Quantum modification of the microscopic reversibility given by $\log \Upsilon$ with respect to the overall coherence $|\alpha|^2_{\rm tot}$ and for different values of the mean thermal photon number $n_{\rm th}$. The experimental data verify that the modification $\log \Upsilon$ scales linearly by increasing the coherence as predicted in Eq.~\eqref{Eq:Modification_Pred}. 
% The dashed line presents the lowest order approximation of the modification up to $(\beta \hbar \omega_0)^2$.
}
\label{Fig:ExpModification1}
\end{figure}
Figure ~\ref{Fig:ExpModification1} shows experimental data and theoretical predictions for the linear scaling of $\log \Upsilon$ with respect to the overall coherence $|\alpha|^2_{\rm tot}$ for different mean numbers of thermal photons.
\begin{figure}[t]
\includegraphics[width=.95\linewidth]{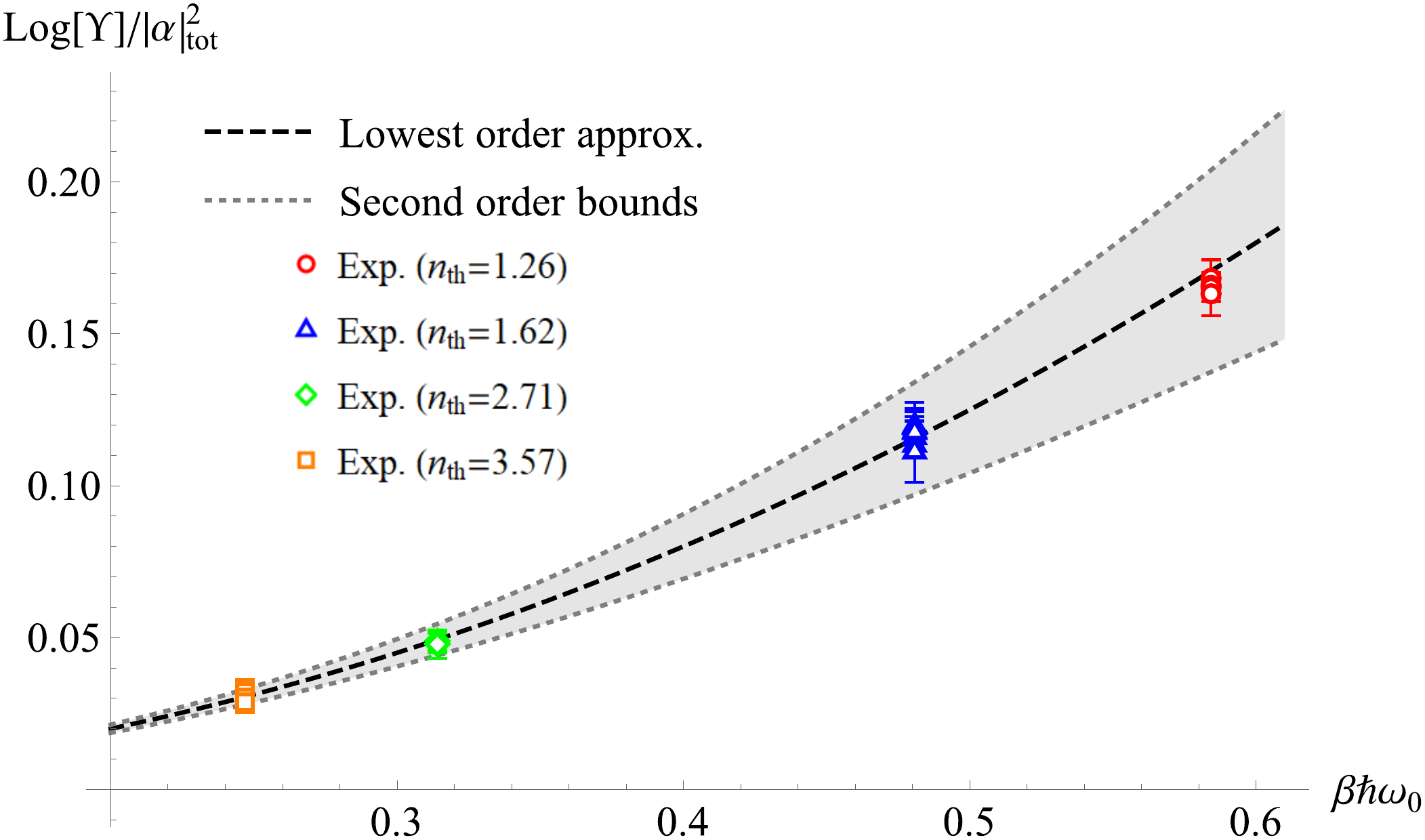}
\caption{ 
Temperature dependence of the quantum correction. The experimental data show the quadratic scaling of $\log \Upsilon / |\alpha|^2_{\rm tot}$ with the inverse temperature $\beta\hbar \omega_0$, regardless of the initial and final coherent state amplitudes. Quantum to classical transition is observed in the high temperature limit $\beta \hbar\omega_0 \ll 1$.
}
\label{Fig:ExpModification2}
\end{figure}
The significant difference between the classical and quantum descriptions is investigated by studying the modification factor $\Upsilon$ in Eq.~\eqref{Eq:Modification_theory} as a function of $|\alpha|^2_{\rm tot}$. The lowest order approximation of the quantum correction in Eq.~\eqref{Eq:Modification_Pred} is given by $\log \Upsilon \approx \frac{(\beta \hbar \omega_0)^2}{2} |\alpha|^2_{\rm tot}$, while the second order bound is obtained as $\left| \log \Upsilon - \frac{(\beta \hbar \omega_0)^2}{2} |\alpha|^2_{\rm tot} \right| \leq \frac{(\beta\hbar\omega_0)^3}{6}|\alpha|^2_{\rm tot} + {\cal O}((\beta \hbar \omega_0)^4)$ by noting that $\left| \Delta |\alpha|^2 \right| \leq |\alpha|^2_{\rm tot}$.

In the low-temperature limit, the quantum modification in the experiment is observed to reach $\Upsilon > 200$ for the transition from $\alpha_i = 2.4$ to $\alpha_f = 5.2$ with $n_{\rm th}=1.26$ ($\beta \hbar \omega_0 \approx 0.58$). This can be understood from the fact that the Gibbs rescaling for the reverse process severely deforms quantum states having coherences between the energy eigenstates when $\beta \hbar \omega_0 \sim 1$. Instead, the quantum-to-classical transition is seen in the high-temperature limit ($\beta \hbar \omega_0 \ll 1$), where the quantum fluctuation relation resembles the classical fluctuation relation in Eq.~\eqref{Eq:ClassicalDB}. For example, by taking $n_{\rm th}=3.57$ (corresponding to $\beta \hbar \omega_0 \approx 0.25$), the modification measured in the experiment is $\Upsilon \approx 1.57$ for $\alpha_i= 3.14$ and $\alpha_f= 2.17$. Figure~\ref{Fig:ExpModification2} clearly shows the quadratic scaling of the quantum correction with respect to the inverse temperature $\beta$, where the leading coefficient after normalization is given by $\frac{\log \Upsilon}{|\alpha|^2_{\rm tot}} \approx \frac{(\beta \hbar \omega_0)^2}{2}$.

%\section{conclusion}
{\it Remarks.---}
We demonstrate the principle of microscopic reversibility in the quantum regime via quantum optics experiments. Our method provides a general extension of the classical thermodynamic interactions to a quantum scenario, by replacing classical phase space points with coherent states in quantum optics and by modeling the system-bath interaction using an energy-preserving unitary operation. The experimental results clearly show that the ratio between the forward and backward quantum thermodynamic processes obeys the quantum description of microscopic reversibility rather than the classical prediction. In particular, we observe that quantum coherences involved in both initial and final states contribute to the modification of microscopic reversibility, especially in the low-temperature limit. Our results show that the quantum modification can aid the symmetry relation between the forward and backward dynamics to understand classical and quantum fluctuation theorems in a single framework, which can be applied to a wide range of physical processes, for example, exploring the Markovian master equation derived by a set of beam-splitters \cite{Kim95}.

\begin{acknowledgments}
 N.B., S.F., M.B., and A.Z. gratefully acknowledge the support of the EU under the ERA-NET QuantERA project ``ShoQC” (Grant No. 731473) and the FET Flagship on Quantum Technologies project ``Qombs” (Grant no. 820419). H.K. is supported by the KIAS Individual Grant No. CG085301 at Korea Institute for Advanced Study. M.S.K. acknowledges the KIST Open Research Program.
\end{acknowledgments}

\newpage
\widetext
\section{Supplementary Material}
\subsection{Direct derivation of the quantum microscopic reversibility condition}
The symmetry relation between the forward and backward transition probabilities of quantum thermodynamic processes has been discussed in Refs.~\cite{Aberg18, Kwon19}. Here, we directly derive an explicit form of the quantum microscopic reversibility condition for the transition between coherent states,
\begin{equation}
\frac{{\cal P}_\rightarrow(\alpha_f | \alpha_i)}{{\cal P}_\leftarrow( \tilde\alpha_i | \tilde\alpha_f)} 
= \exp\left[-\frac{|\alpha_f|^2}{n_{\rm th} +1} + \frac{|\alpha_i|^2}{n_{\rm th}}\right],
\end{equation}
where $\tilde\alpha_i = \bar\alpha_i e^{\beta \hbar \omega_0/2}$ and $\tilde\alpha_f =  \bar\alpha_f e^{-\beta \hbar \omega_0/2}$.
\begin{proof}
We first note that a Gibbs-rescaled coherent state can be expressed as the following form:
$$
\ket{\alpha e^{\pm \beta \hbar \omega_0/2}} = {\cal N}_\pm (\alpha) e^{\pm \beta \hat{H}/2} \ket{\alpha},
$$
where $\hat{H} = \hbar \omega_0 \left( \hat{a}^\dagger \hat{a} + \frac{1}{2} \right)$ and ${\cal N}_\pm(\alpha) = \left( \bra\alpha e^{\pm \beta \hat{H}} \ket{\alpha} \right)^{-\frac{1}{2}}$. Using the Gibbs-rescaled coherent states, we express the backward transition probability as
\begin{equation}
\begin{aligned}
{\cal P}_\leftarrow(\tilde\alpha_i | \tilde\alpha_f)
&= {\Tr} \left[ \hat{U}_R \left(\ket{\tilde\alpha_f} \bra{\tilde\alpha_f} \otimes \hat{\rho}_{\rm th}\right) \hat{U}_R^\dagger \left( \ket{\tilde\alpha_i} \bra{\tilde\alpha_i} \otimes \mathbb{\hat{1}}_B \right) \right]\\
&= {\cal N}_+^2(\alpha_i) {\cal N}_-^2(\alpha_f) {\Tr} \left[ \hat{U}_R \left(e^{-\beta \hat{H}/2} \ket{\bar\alpha_f} \bra{\bar\alpha_f} e^{-\beta \hat{H}/2} \otimes \hat{\rho}_{\rm th} \right) \hat{U}_R^\dagger \left( e^{\beta \hat{H}/2}\ket{\bar\alpha_i} \bra{\bar\alpha_i} e^{\beta \hat{H}/2} \otimes \mathbb{\hat{1}}_B \right) \right].
\end{aligned}
\label{eq:supp_p_rev}
\end{equation}
From the energy conservation condition $[ \hat{H}_{I}, \hat{H} + \hat{H}_B] = 0$ and time-reversal symmetry of the system and bath Hamiltonians, $H(\hat{x},\hat{p}) = H(\hat{x}, - \hat{p})$ and $H_B(\hat{x}_B,\hat{p}_B) = H_B(\hat{x}_B, - \hat{p}_B)$, we observe that $[\hat{H}_I^R, \hat{H} + \hat{H}_B] = 0 \Rightarrow [\hat{U}_R, \hat{H} + \hat{H}_B] =0 $. We then express the quantum state following the reverse dynamics as
$$
\begin{aligned}
\hat{U}_R \left( e^{-\beta \hat{H}/2} \ket{\bar\alpha_f} \bra{\bar\alpha_f} e^{-\beta \hat{H}/2} \otimes \hat{\rho}_{\rm th} \right) \hat{U}_R^\dagger &= 
\frac{1}{Z_B} \hat{U}_R \left( e^{-\beta \hat{H}/2} \otimes e^{-\beta \hat{H}_B/2} \right) \left(\ket{\bar\alpha_f} \bra{\bar\alpha_f} \otimes \mathbb{1}_B \right) \left( e^{-\beta \hat{H}/2} \otimes e^{-\beta \hat{H}_B/2} \right) \hat{U}_R^\dagger\\
&= \frac{1}{Z_B} \left( e^{-\beta \hat{H}/2} \otimes e^{-\beta \hat{H}_B/2} \right) \hat{U}_R \left( \ket{\bar\alpha_f} \bra{\bar\alpha_f} \otimes \mathbb{1}_B \right) \hat{U}_R^\dagger \left( e^{-\beta \hat{H}/2} \otimes e^{-\beta \hat{H}_B/2} \right),
\end{aligned}
$$
where $\hat\rho_{\rm th} = \frac{e^{-\beta \hat{H}_B}}{Z_B}$ with $Z_B = {\rm Tr}\left[ e^{-\beta \hat{H}_B} \right]$. By substituting this formula to Eq.~\eqref{eq:supp_p_rev}, we obtain
\begin{equation}
\begin{aligned}
&{\cal P}_\leftarrow(\tilde\alpha_i | \tilde\alpha_f) \\
&\quad = \frac{{\cal N}_+^2(\alpha_i) {\cal N}_-^2(\alpha_f)}{Z_B}
{\rm Tr}\left[ \left( e^{-\beta \hat{H}/2} \otimes e^{-\beta \hat{H}_B/2} \right) \hat{U}_R \left(\ket{\bar\alpha_f} \bra{\bar\alpha_f} \otimes \mathbb{1}_B \right) \hat{U}_R^\dagger \left(e^{-\beta \hat{H}/2} \otimes e^{-\beta \hat{H}_B/2} \right) \left( e^{\beta \hat{H}/2}\ket{\bar\alpha_i} \bra{\bar\alpha_i} e^{\beta \hat{H}/2} \otimes \mathbb{\hat{1}}_B \right) \right]\\
&\quad = \frac{{\cal N}_+^2(\alpha_i) {\cal N}_-^2(\alpha_f)}{Z_B}
{\rm Tr}\left[ \hat{U}_R^\dagger \left(e^{-\beta \hat{H}/2} \otimes e^{-\beta \hat{H}_B/2} \right) \left( e^{\beta \hat{H}/2}\ket{\bar\alpha_i} \bra{\bar\alpha_i} e^{\beta \hat{H}/2} \otimes \mathbb{\hat{1}}_B \right) \left(e^{-\beta \hat{H}/2} \otimes e^{-\beta \hat{H}_B/2} \right) \hat{U}_R (\ket{\bar\alpha_f} \bra{\bar\alpha_f} \otimes \mathbb{1}_B ) \right]\\
&\quad = \frac{{\cal N}_+^2(\alpha_i) {\cal N}_-^2(\alpha_f)}{Z_B}
{\rm Tr}\left[ \hat{U}_R^\dagger \left( \ket{\bar\alpha_i} \bra{\bar\alpha_i} \otimes e^{-\beta H_B}  \right) \hat{U}_R (\ket{\bar\alpha_f} \bra{\bar\alpha_f} \otimes \mathbb{1}_B ) \right]\\
&\quad = {\cal N}_+^2(\alpha_i) {\cal N}_-^2(\alpha_f)
{\rm Tr}\left[ \hat{U}_R^\dagger \left( \ket{\bar\alpha_i} \bra{\bar\alpha_i} \otimes \hat{\rho}_{\rm th} \right) \hat{U}_R \left(\ket{\bar\alpha_f} \bra{\bar\alpha_f} \otimes \mathbb{1}_B \right) \right].
%&\quad = {\cal N}_+^2(\alpha_i) {\cal N}_-^2(\alpha_f) {\cal P}_{\rightarrow}(\bar\alpha_f | \bar\alpha_i).
\end{aligned}
\label{Eq:supp_p_fr}
\end{equation}
As both the thermal state and the identity operator are invariant under time-reversal, $(\hat{x}_B, \hat{p}_B) \rightarrow (\hat{x}_B, -\hat{p}_B)$, there exist the representations $\hat{\rho}_{\rm th} = \sum_\lambda P_{\hat{\rho}_{\rm th}}(\lambda) \ket{\lambda}\bra{\lambda} = \sum_\lambda P_{\hat{\rho}_{\rm th}}(\lambda) \ket{\bar\lambda}\bra{\bar\lambda}$ and $\mathbb{1}_B = \sum_\lambda P_{\mathbb{1}_B}(\lambda) \ket{\bar\lambda}\bra{\bar\lambda}$, where $\bar{\lambda}$ denotes time-reversal of the bath coordinates. This observation leads to 
\begin{equation}
\begin{aligned}
{\rm Tr}\left[ \hat{U}_R^\dagger \left( \ket{\bar\alpha_i} \bra{\bar\alpha_i} \otimes \hat{\rho}_{\rm th} \right) \hat{U}_R \left(\ket{\bar\alpha_f} \bra{\bar\alpha_f} \otimes \mathbb{1}_B \right) \right] 
&= \sum_{\lambda, \lambda'} P_{\hat{\rho}_{\rm th}}(\lambda) P_{\mathbb{1}_B}(\lambda') {\rm Tr}\left[ \hat{U}_R^\dagger \left( \ket{\bar\alpha_i} \bra{\bar\alpha_i} \otimes \ket{\bar\lambda} \bra{\bar\lambda} \right) \hat{U}_R \left(\ket{\bar\alpha_f} \bra{\bar\alpha_f} \otimes \ket{\bar\lambda'}\bra{\bar\lambda'} \right) \right]\\
&= \sum_{\lambda, \lambda'} P_{\hat{\rho}_{\rm th}}(\lambda) P_{\mathbb{1}_B}(\lambda') 
\bra{\bar\alpha_f, \bar\lambda'} \hat{U}_R^\dagger \ket{\bar{\alpha}_i, \bar\lambda}
\bra{\bar\alpha_i, \bar\lambda} \hat{U}_R \ket{\bar{\alpha}_f, \bar\lambda'} \\
&= \sum_{\lambda, \lambda'} P_{\hat{\rho}_{\rm th}}(\lambda) P_{\mathbb{1}_B}(\lambda') 
\bra{\alpha_i, \lambda} \hat{U}^\dagger \ket{\alpha_f, \lambda'}
\bra{\alpha_f, \lambda'} \hat{U} \ket{\alpha_i, \lambda} \\
&= {\rm Tr} \left[ \hat{U} \left( \ket{\alpha_i}\bra{\alpha_i} \otimes \hat{\rho}_{\rm th}\right) \hat{U}^\dagger \left( \ket{\alpha_f}\bra{\alpha_f} \otimes \mathbb{1}_B \right) \right] \\
&= {\cal P}_{\rightarrow}(\alpha_f|\alpha_i).
\end{aligned}
\label{Eq:supp_p_t_rev}
\end{equation}
Here, the third equality is obtained from the time-reversal symmetry of the quantum dynamics $\left| \bra{\bar\alpha_i, \bar\lambda} \hat{U}_R \ket{\bar{\alpha}_f, \bar\lambda'} \right|^2 = \left| \bra{\alpha_f, \lambda'} \hat{U} \ket{\alpha_i, \lambda} \right|^2$. Finally, by rearranging the expressions in Eqs.~\eqref{Eq:supp_p_fr} and \eqref{Eq:supp_p_t_rev}, we obtain
\begin{equation}
\frac{{\cal P}_{\rightarrow}(\alpha_f | \alpha_i)}{{\cal P}_\leftarrow(\tilde\alpha_i | \tilde\alpha_f)} = \frac{1}{{\cal N}_+^2(\alpha_i) {\cal N}_-^2(\alpha_f)} = \bra{\alpha_i} e^{\beta \hat{H}} \ket{\alpha_i} \bra{\alpha_f} e^{-\beta \hat{H}} \ket{\alpha_f}.
\label{eq:supp_QDB}
\end{equation}
An explicit form of the right-hand side of the equation can be evaluated as
$$
\begin{aligned}
\bra{\alpha_i} e^{\beta \hat{H}} \ket{\alpha_i} 
&= e^{-|\alpha_i|^2(1-\exp[\beta\hbar\omega_0])} = e^{\frac{|\alpha_i|^2}{n_{\rm th}}} \\
\bra{\alpha_f} e^{- \beta \hat{H}} \ket{\alpha_f} &= e^{-|\alpha_f|^2(1-\exp[-\beta\hbar\omega_0])} = e^{-\frac{|\alpha_f|^2}{n_{\rm th}+1}},
\end{aligned}
$$
where $\frac{1}{n_{\rm th}} = e^{\beta \hbar\omega_0} -1 $ and $\frac{1}{n_{\rm th}+1} = 1 - e^{-\beta \hbar\omega_0}$ by defining $n_{\rm th} = \frac{e^{-\beta\hbar\omega_0}}{1-e^{-\beta\hbar\omega_0}}$. By noting that $Q = \Delta E = \bra{\alpha_f} \hat{H} \ket{\alpha_f} - \bra{\alpha_i} \hat{H} \ket{\alpha_i}$, we derive a quantum correction,
$$
\frac{{\cal P}_\rightarrow(\alpha_f | \alpha_i)}{{\cal P}_\leftarrow( \tilde\alpha_i | \tilde\alpha_f)} 
= \Upsilon e^{- \beta Q},
$$
with
$$
\begin{aligned}
\Upsilon &= \frac{\bra{\alpha_i}e^{\beta \hat{H}} \ket{\alpha_i}}{e^{\beta \bra{\alpha_i} \hat{H}\ket{\alpha_i}}} \frac{\bra{\alpha_f}e^{-\beta \hat{H}} \ket{\alpha_f}}{e^{-\beta \bra{\alpha_f} \hat{H}\ket{\alpha_f}}}\\
&=\exp\left[{-|\alpha_i|^2 (1-e^{\beta\hbar\omega_0} + \beta\hbar\omega_0) - |\alpha_f|^2 (1-e^{-\beta\hbar\omega_0}-\beta\hbar\omega_0)} \right]\\
&= \exp[(\cosh(\beta \hbar \omega_0) -1) (|\alpha_i|^2 + |\alpha_f|^2 ) - (\sinh(\beta \hbar \omega_0) -\beta\hbar\omega_0)\left( |\alpha_f|^2 - |\alpha_i|^2 \right)].
\end{aligned}
$$
\end{proof}

\subsection{Experimental setup}
\begin{figure}[h]
	\begin{center}
		\includegraphics[width=.75\linewidth]{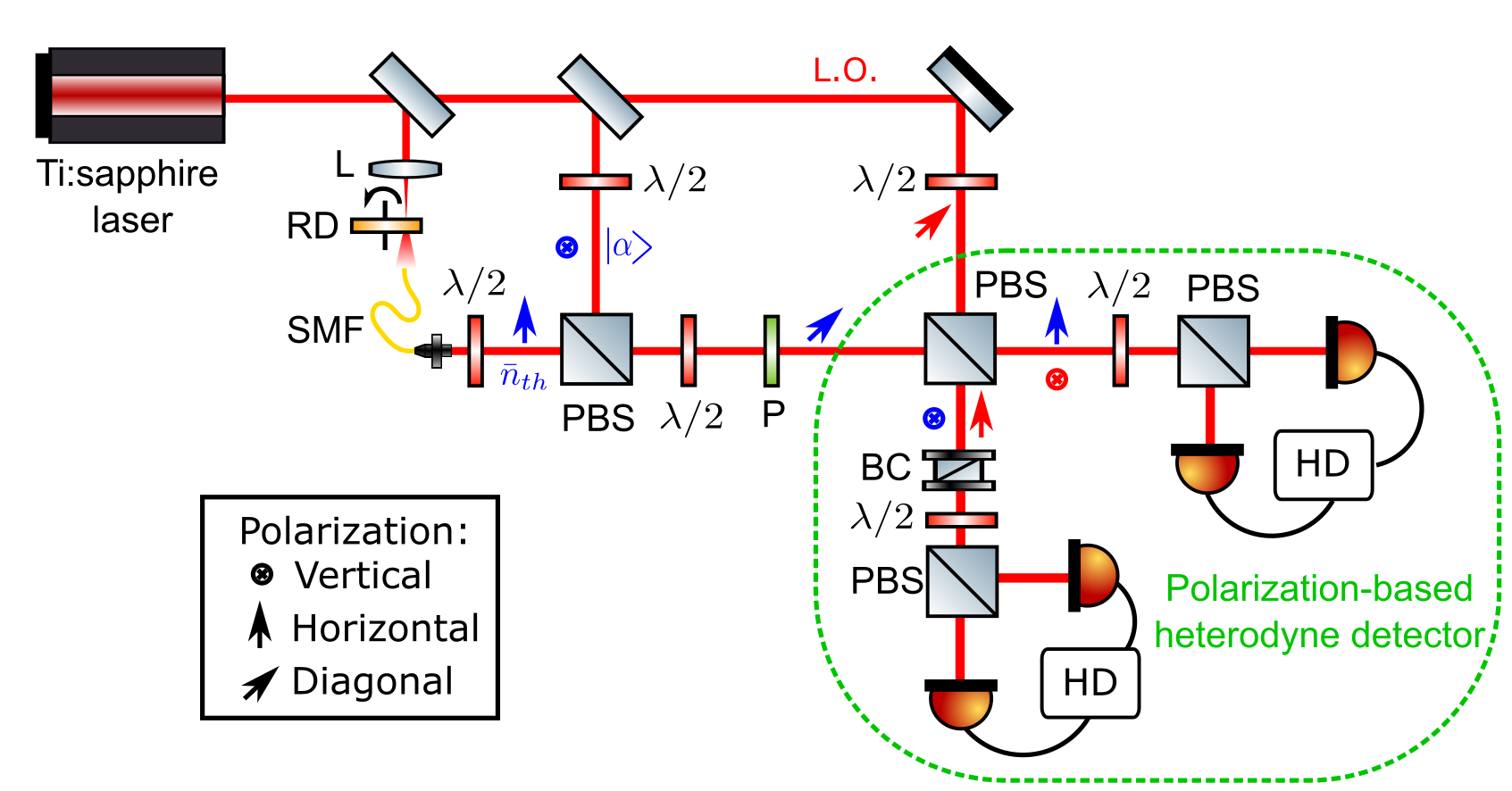}
		\caption{Scheme of the experimental setup. The box explains the symbols employed to indicate the polarization direction. Red arrows refer to LO, blue arrows refer to signal. Abbreviations: L: lens, RD: rotating disk, $\lambda /2$: half-wave plate, PBS: polarizing beam-splitter, P: polarizer, BC: Babinet compensator, HD: homodyne detector. }
		\label{fig:Setup}
	\end{center}
\end{figure}

The experimental setup, shown in Figure \ref{fig:Setup}, is based on a mode-locked Ti:Sa laser emitting 1.5 ps-long pulses at 786 nm with a repetition rate of 81 MHz. Its output is split into three parts. The first one provides the local oscillator (LO) pulses for heterodyne detection. The second part is employed to produce a pseudo-thermal state of light. A lens (L) focuses the beam onto a rotating ground glass disk (RD) \cite{Arecchi1965}. Collecting a small portion of the light scattered by the disk into a single-mode fiber, we obtain a thermal state in a single spatial mode \cite{Zavatta2007, Parigi2009}. 
The third beam gives the coherent states. Two variable attenuators (not shown in the figure), placed along the thermal and coherent state paths, are used to control the thermal mean photon number $\bar{n}_{\rm th}$ and the coherent state amplitude $\alpha$.

Thermal and coherent states are mixed in a polarization-based beam-splitter with variable reflectivity composed as follows. The thermal states are vertically polarized while the coherent states are horizontally polarized; their spatial modes are mixed via a polarizing beam-splitter (PBS). Then, a half-wave plate (HWP) rotates their polarization before entering into a polarizer (P), diagonally oriented. The combination of these three devices plays the role of a beam-splitter with adjustable reflectivity, which allows us to control the ratio of coherent and thermal states simulating the thermodynamic process.

Throughout the experiment, the reflectivity of the beam-splitter mixing coherent and thermal states has been set to two different values: 85\% and 70\%, depending on the coherent state amplitudes and mean thermal photon numbers used. This is due to the need of a sufficiently large mean field amplitude of the states reaching the two homodyne detectors, for a reliable lock of the relative signal-LO phase and in order to remain in the detectors' dynamic range. Experimental results confirm that the ${\cal P}_\rightarrow/{\cal P}_\leftarrow$ ratio is independent on the beam-splitter reflectivity.

The detection is performed via a polarization-based heterodyne detector. The signal, composed by the mix of thermal and coherent states, and the LO enter the central PBS with diagonal polarizations. Therefore, they are equally split between the two arms of the heterodyne. In each arm a HWP, which rotates the polarization back to diagonal, and a second PBS play the role of the balanced beam-splitter of the homodyne detector (HD). 
A crucial parameter of the heterodyne detector is the phase relationship between the quadrature measured by HD1 and the one measured by HD2, which must be exactly $\pi/2$. For this reason we inserted a Babinet compensator (BC) in the path to HD1, which allows us to fine tune the delay between signal and LO.

We first use coherent states with real values of $\alpha$. This simplifies the procedure since the Gibbs rescaled $\tilde \alpha$ remains real and thus all the $Q$-functions are centred on the $\hat{x}$ quadrature axis. Then, we also test the use of coherent states with complex amplitudes (data points corresponding to a mean thermal photon number $n_{th}=1.22$ in Fig.~3 of the main text). In such a case, the Gibbs rescaled values are complex conjugated and thus the $Q$-functions of the forward and time-reversed process are located in two different quadrants of the phase space, as shown in Fig.~3 of the main text. For the sake of simplicity, when starting from an initial coherent state with complex amplitude $\alpha_i=|\alpha| \exp(i\phi)$, we evaluate the $Q$-function at values of $\alpha_f$ having the same phase of $\alpha_i$ ($\theta_i=\theta_f=\phi$). This simplifies the experimental procedure since the LO phase has to be locked only to two values: $\theta_i=\phi$ for the forward process, and $\tilde\theta_f=-\phi$ for the time-reversed one.

$\hat{x}$ and $\hat{p}$ quadrature data are simultaneously collected from the two HDs, with the global LO phase locked via a piezo-mounted mirror. These data directly provide the Husimi $Q$-function \cite{Richter98}, whose distribution is estimated via a maximum likelihood algorithm after 50000 $\hat{x}$ and $\hat{p}$ quadrature measurements. The confidence intervals of the distribution parameters are calculated using a bootstrap method: 1000 random values are sampled from the distribution and are employed to evaluate again the distribution parameters; the procedure is repeated 1000 times and the confidence intervals are retrieved from the standard deviation of the parameters.


\begin{thebibliography}{99}
\bibitem{Gallavotti95} G. Gallavotti and E. G. D. Cohen, \textit{Dynamical Ensembles in Nonequilibrium Statistical Mechanics}, Phys. Rev. Lett. \textbf{74}, 2694 (1995).
\bibitem{Jarzynski97} C. Jarzynski, \textit{Nonequilibrium Equality for Free Energy Differences}, Phys. Rev. Lett. \textbf{78}, 2690 (1997).
\bibitem{Crooks99} G. E. Crooks, \textit{Entropy production fluctuation theorem and the nonequilibrium work relation for free energy differences}, Phys. Rev. E \textbf{60}, 2721 (1999).
\bibitem{Liphardt02} J. Liphardt, S. Dumont, S. B. Smith, I. Tinoco, Jr., and C. Bustamante, \textit{Equilibrium Information from Nonequilibrium Measurements in an Experimental Test of Jarzynski’s Equality}, Science \textbf{296}, 1832 (2002).
\bibitem{Collin05} D. Collin, F. Ritort, C. Jarzynski, S. B. Smith, I. Tinoco, Jr., and C. Bustamante, \textit{Verification of the Crooks Fluctuation Theorem and Recovery of RNA Folding Free Energies}, Nature (London) \textbf{437}, 231 (2005).
\bibitem{Blickle06} V. Blickle, T. Speck, L. Helden, U. Seifert, and C. Bechinger, \textit{Thermodynamics of a Colloidal Particle in a Time-Dependent Nonharmonic Potential}, Phys. Rev. Lett. \textbf{96}, 070603 (2006).
\bibitem{Saira12} O.-P. Saira, Y. Yoon, T. Tanttu, M. M\"ott\"onen, D. V. Averin, and J. P. Pekola, \textit{Test of the Jarzynski and Crooks Fluctuation Relations in an Electronic System}, Phys. Rev. Lett. \textbf{109}, 180601 (2012).
\bibitem{Esposito09} M. Esposito, U. Harbola, and S. Mukamel, \textit{Nonequilibrium Fluctuations, Fluctuation Theorems, and Counting Statistics in Quantum Systems}, Rev. Mod. Phys. \textbf{81}, 1665 (2009).
\bibitem{Campisi11} M. Campisi, P. H{\"a}nggi, and P. Talkner, \textit{Colloquium: Quantum Fluctuation Relations: Foundations and Applications}, Rev. Mod. Phys. \textbf{83}, 771 (2011).
\bibitem{Deffner11} S. Deffner, M. Brunner and E. Lutz, \textit{Quantum fluctuation theorems in the strong damping limit}, Europhys. Lett. \textbf{94}, 30001 (2011).
\bibitem{Deffner13} S. Deffner, \textit{Quantum Entropy Production in Phase Space}, Europhys. Lett. \textbf{103}, 30001 (2013).
\bibitem{Hanggi15} P. H{\"a}nggi and P. Talkner, \textit{The Other QFT}, Nat. Phys. \textbf{11}, 108 (2015).
\bibitem{Jarzynski15} C. Jarzynski, H. T. Quan, and S. Rahav, \textit{Quantum-Classical Correspondence Principle for Work Distributions}, Phys. Rev. X \textbf{5}, 031038 (2015).
\bibitem{Funo18} K. Funo, M. Ueda, and T. Sagawa, \textit{Quantum Fluctuation Theorems} in \textit{Thermodynamics in the Quantum Regime (Springer, 2018)} pp. 249–273.
\bibitem{Taranto18} P. Taranto, K. Modi, and F. A. Pollock, \textit{Emergence of a fluctuation relation for heat in nonequilibrium Landauer processes}, Phys. Rev. E \textbf{97}, 052111 (2018).


\bibitem{Goold16} J. Goold, M. Huber, A. Riera, L. del Rio, and P. Skrzypczyk, \textit{The Role of Quantum Information in Thermodynamics: A Topical Review}, J. Phys. A \textbf{49}, 143001 (2016).
\bibitem{Alhambra16} {\'A}. M. Alhambra, L. Masanes, J. Oppenheim, and C. Perry, \textit{Fluctuating Work: From Quantum Thermodynamical Identities to a Second Law Equality}, Phys. Rev. X \textbf{6}, 041017 (2016).
\bibitem{Binder19} F. Binder, L. A. Correa, C. Gogolin, J. Anders, and G. Adesso, eds., \textit{Thermodynamics in the Quantum Regime} (Springer International, 2019).
\bibitem{Landi21} G. T. Landi and M. Paternostro, \textit{Irreversible entropy production: From classical to quantum}, Rev. Mod. Phys. \textbf{93}, 035008 (2021).

\bibitem{Talkner07} P. Talkner, E. Lutz, and P. H{\"a}nggi, \textit{Fluctuation Theorems: Work Is Not an Observable}, Phys. Rev. E \textbf{75}, 050102(R) (2007).
\bibitem{Bera19} M. N. Bera, A. Riera, M. Lewenstein, Z. B. Khanian and A. Winter, \textit{Thermodynamics as a Consequence of Information Conservation}, Quantum \textbf{3}, 121 (2019).
\bibitem{Scully03} M. O. Scully, M. S. Zubairy, G. S. Agarwal, and H. Walther, \textit{Extracting Work from a Single Heat Bath via Vanishing Quantum Coherence}, Science \textbf{299}, 862 (2003).
\bibitem{Pikovski12} I. Pikovski, M. R. Vanner, M. Aspelmeyer, M. S. Kim, and {\v C}. Brukner , \textit{Probing Planck-scale physics with quantum optics} Nat. Phys. \textbf{8}, 393 (2012).
\bibitem{Vidrighin16} M. D. Vidrighin, O. Dahlsten, M. Barbieri, M. S. Kim, V. Vedral, and I. A. Walmsley, \textit{Photonic Maxwell’s Demon}, Phys. Rev. Lett. \textbf{116}, 050401 (2016).



\bibitem{Albash13} T. Albash, D. A. Lidar, M. Marvian, and P. Zanardi, \textit{Fluctuation theorems for quantum processes}, Phys. Rev. E \textbf{88}, 032146 (2013).
\bibitem{Batalhao14} T. B. Batalh\~{a}o, A. M. Souza, L. Mazzola, R. Auccaise, R. S. Sarthour, I. S. Oliveira, J. Goold, G. De Chiara, M. Paternostro, and R. M. Serra, \textit{Experimental Reconstruction of Work Distribution and Study of Fluctuation Relations in a Closed Quantum System}, Phys. Rev. Lett. \textbf{113}, 140601 (2014).
\bibitem{Aberg18} J. {\AA}berg, \textit{Fully Quantum Fluctuation Theorems}, Phys. Rev. X \textbf{8}, 011019 (2018).
\bibitem{Manzano18} G. Manzano, J. M. Horowitz, and J. M. R. Parrondo, \textit{Quantum Fluctuation Theorems for Arbitrary Environments: Adiabatic and Nonadiabatic Entropy Production}, Phys. Rev. X \textbf{8}, 031037 (2018).
\bibitem{Holmes19} Z. Holmes, S. Weidt, D. Jennings, J. Anders, and F. Mintert, \textit{Coherent Fluctuation Relations: From the Abstract to the Concrete}, Quantum \textbf{3}, 124 (2019).
\bibitem{Kwon19} H. Kwon and M. S. Kim, \textit{Fluctuation theorems for a quantum channel}, Phys. Rev. X \textbf{9}, 031029 (2019).
\bibitem{Micadei20} K. Micadei, G. T. Landi, and E. Lutz, \textit{Quantum Fluctuation Theorems beyond Two-Point Measurements}, Phys. Rev. Lett. \textbf{124}, 090602 (2020).
\bibitem{Micadei21} K. Micadei, J. P.~S. Peterson, A. M. Souza, R. S. Sarthour, I. S. Oliveira, G. T. Landi, R. M. Serra, and E. Lutz, \textit{Experimental Validation of Fully Quantum Fluctuation Theorems Using Dynamic Bayesian Networks} Phys. Rev. Lett. \textbf{127}, 180603 (2021).
\bibitem{Goold21} J. Goold and  K. Modi, \textit{Fluctuation theorem for nonunital dynamics} AVS Quantum Sci. \textbf{3}, 045001 (2021).
\bibitem{Yada22} T. Yada, N. Yoshioka, and T. Sagawa, \textit{Quantum Fluctuation Theorem under Quantum Jumps with Continuous Measurement and Feedback}, Phys. Rev. Lett. 128, 170601 (2022).


\bibitem{Zurek93} W. H. Zurek, S. Habib, and J. P. Paz, \textit{Coherent states via decoherence}, Phys. Rev. Lett. \textbf{70}, 1187 (1993).
\bibitem{Mandel86} L. Mandel, \textit{Non-Classical States of the Electromagnetic Field}, Phys. Scr. \textbf{T12}, 34 (1986).

\bibitem{footnote} In the basis of position eigenstate $\ket{x}$ such that $\hat{x}\ket{x} = x\ket{x}$, the time-reversal of the state $\ket\psi = \int_{-\infty}^\infty dx \psi(x) \ket{x}$ becomes $\ket{\bar\psi} = \int_{-\infty}^\infty dx \psi^*(x) \ket{x}$.

\bibitem{Janzing00} D. Janzing, P. Wocjan, R. Zeier, R. Geiss, and T. Beth, \textit{Thermodynamic Cost of Reliability and Low Temperatures: Tightening Landauer's Principle and the Second Law}, Int. J. Theor. Phys. \textbf{39}, 2717 (2000).
\bibitem{Morodecki13} M. Horodecki and J. Oppenheim, \textit{Fundamental limitations for quantum and nanoscale thermodynamics}, Nat. Commun. \textbf{4}, 2059 (2013).
\bibitem{Lostaglio18} M. Lostaglio, {\'A}. M. Alhambra, and C. Perry, \textit{Elementary Thermal Operations}, Quantum \textbf{2}, 52 (2018).
\bibitem{Kwon18} H. Kwon, H. Jeong, D. Jennings, B. Yadin, and M. S. Kim, \textit{Clock--Work Trade-Off Relation for Coherence in Quantum Thermodynamics}, Phys. Rev. Lett. \textbf{120}, 150602 (2018).
\bibitem{Richter98} Th. Richter, \textit{Determination of photon statistics and density matrix from double homodyne detection measurements}, Journal of Modern Optics, \textbf{45:8}, 1735-1749 (1998).
\bibitem{Arecchi1965} F.T. Arecchi, \textit{Measurement of the Statistical Distribution of Gaussian and Laser Sources}, Phys. Rev. Lett. \textbf{15}, 912 (1965).
\bibitem{Zavatta2007} A. Zavatta, V. Parigi, and M. Bellini, \textit{Experimental nonclassicality of single-photon-added thermal light states}, Phys. Rev. A \textbf{75}, 052106 (2007).
\bibitem{Parigi2009} V. Parigi, A. Zavatta, and M. Bellini, \textit{Implementation of single-photon creation and annihilation operators: experimental issues in their application to thermal states of light}, J. Phys. B: At. Mol. Opt. Phys \textbf{42}, 114005 (2009). 
\bibitem{Suppl} Supplemental Material for detailed derivation and experimental setup.
\bibitem{Kim95} M. S. Kim and N. Imoto, \textit{Phase-sensitive reservoir modeled by beam splitters}, Phys. Rev. A \textbf{52}, 2401 (1995).

\end{thebibliography}
\end{document}